\documentclass[9pt,conference]{IEEEtran}
\IEEEoverridecommandlockouts

\usepackage{cite}
\usepackage{amsmath,amssymb,amsfonts}
\usepackage{multicol}
\usepackage{amsthm}
\usepackage{graphicx}
\usepackage{subcaption}
\usepackage{textcomp}
\usepackage{xcolor}
\usepackage{booktabs}
\usepackage{comment}
\usepackage{tabularx}
\usepackage{multirow}
\usepackage{bm}
\usepackage{url}
\usepackage{xspace}
\usepackage{pifont}
\usepackage{arydshln}
\usepackage{verbatim}
\usepackage[referable]{threeparttablex}
\usepackage{calc}
\usepackage[normalem]{ulem}
\usepackage{color}
\usepackage{algorithm}
\usepackage{algpseudocode}
\usepackage{tikz}
\usepackage{flushend}
\usepackage{xcolor}
\usepackage{colortbl}
\usetikzlibrary{shapes.geometric, arrows}
\usepackage[capitalise]{cleveref}
\tikzstyle{startstop} = [rectangle, rounded corners, minimum width=3cm, minimum height=1cm,text centered, draw=black]
\tikzstyle{io} = [trapezium, trapezium left angle=70, trapezium right angle=110, minimum width=3cm, minimum height=1cm, text centered, draw=black]
\tikzstyle{process} = [rectangle, minimum width=3cm, minimum height=1cm, text centered, draw=black]
\tikzstyle{arrow} = [thick,->,>=stealth]

\newcommand{\AUTOPDR}{\texttt{AutoPDR}\xspace}

\newcommand{\blackcircled}[1]{%
  \tikz[baseline=(char.base)]{%
    \node[shape=circle,draw,inner sep=1pt,fill=black] (char) {\textcolor{white}{\scriptsize #1}};%
  }%
}

\def\BibTeX{{\rm B\kern-.05em{\sc i\kern-.025em b}\kern-.08em
    T\kern-.1667em\lower.7ex\hbox{E}\kern-.125emX}}
\begin{document}

\title{\AUTOPDR: Circuit-Aware Solver Configuration Prediction for
Hardware Model Checking\\
\thanks{}
}

\author{
\IEEEauthorblockN{
Guangyu Hu\IEEEauthorrefmark{1},
Chen Chen\IEEEauthorrefmark{2},
Xiaofeng Zhou\IEEEauthorrefmark{1},
Jiaxi Zhang\IEEEauthorrefmark{3},
Wei Zhang\IEEEauthorrefmark{1},
Hongce Zhang\IEEEauthorrefmark{2}
}
\IEEEauthorblockA{\IEEEauthorrefmark{1}The Hong Kong University of Science and Technology, \{ghuae, xzhoubu\}@connect.ust.hk, wei.zhang@ust.hk}
\IEEEauthorblockA{\IEEEauthorrefmark{2}The Hong Kong University of Science and Technology (Guangzhou), cchen099@connect.hkust-gz.edu.cn, hongcezh@hkust-gz.edu.cn}
\IEEEauthorblockA{\IEEEauthorrefmark{3}Peking University, zhangjiaxi@pku.edu.cn}
}

\maketitle

\begin{abstract}
Property Directed Reachability (PDR) is a powerful algorithm for formal verification of hardware and software systems, but its performance is highly sensitive to parameter configurations. Manual parameter tuning is time-consuming and requires domain expertise, while traditional automated parameter tuning frameworks are not well-suited for time-sensitive verification tasks like PDR. This paper presents a circuit-aware solver configuration framework that employs graph learning for intelligent heuristic selection in PDR-based verification. Our approach combines graph representations with static circuit features
to predict optimal PDR solving configurations for specific circuits. 
We incorporate expert prior knowledge through constraint-based parameter filtering to eliminate invalid and inefficient configurations and reduce 78\% search space. Our feature extraction pipeline captures structural, functional, and connectivity characteristics of circuit topology and component patterns. Experimental evaluation on a comprehensive benchmark suite demonstrates significant performance improvements compared to default configurations and commonly-used settings. The system successfully identifies circuit-specific parameter patterns and automatically selects the most suitable solving strategies based on circuit characteristics, making it a practical tool for automated formal verification workflows. The code is available at \url{https://github.com/Gy-Hu/AutoPDR}.
\end{abstract}

\begin{IEEEkeywords}
formal verification, property directed reachability, IC3, parameter tuning, machine learning, graph neural networks, automated reasoning, solver optimization.
\end{IEEEkeywords}

\section{Introduction}

Formal verification has become an essential component of modern hardware and software design, providing mathematical guarantees about system correctness. Among the various verification techniques, Property Directed Reachability (PDR), also known as IC3~\cite{bradley2011sat}, has emerged as one of the most effective algorithms for proving safety properties of finite-state systems. PDR's success stems from its ability to incrementally construct inductive invariants while avoiding the explicit enumeration of reachable states, making it particularly well-suited for large-scale verification problems~\cite{hoder2011generalized,een2011inductive}.

However, PDR's effectiveness is highly dependent on its parameter configuration. 
One of the commonly-used PDR implementations is built in \texttt{ABC}~\cite{een2011inductive}, which exposes numerous tunable parameters that control various aspects of its operation, including generalization strategies, abstraction mechanisms, proof obligation handling, and heuristic choices. These parameters significantly impact both the runtime performance and the likelihood of successful verification~\cite{Griggio2015,Cabodi2017}. Unfortunately, finding optimal parameter settings is a challenging task that typically requires extensive domain expertise and time-consuming manual experimentation~\cite{Hoos2012}. Moreover, the combinatorial explosion of parameter combinations makes parallel exploration computationally infeasible, as it would quickly exhaust available computational resources, such as CPU cores and memory.

The parameter tuning problem in PDR is particularly complex due to several factors. First, the optimal parameter configuration is highly dependent on the structural characteristics of the circuit being verified. Parameters that work well for control-intensive designs may perform poorly on datapath-heavy circuits. Second, the parameter space is large and contains complex interdependencies—certain parameter combinations are logically invalid or lead to degenerate behavior. Third, the relationship between circuit features and optimal parameters is intricate and difficult to capture with traditional heuristics. 
These challenges highlight the need for a more systematic and learning-based approach that can capture the intricate relationships between circuit structure and parameter effectiveness.

Current approaches to PDR parameter tuning fall into two categories: manual tuning by experts and simple automated search methods. Manual tuning is labor-intensive, as it requires deep understanding of both the algorithm and the target domain, and does not scale to diverse verification workloads. Automated search methods, such as random sampling or grid search, fail to exploit the structural information available in the verification problem and often waste computational resources exploring invalid or suboptimal configurations~\cite{hutter2016cssc,leventi2021ml_sat}. Furthermore, PDR engines frequently consume substantial memory resources, and solving times often exceed one hour, making such brute-force approaches impractical and unreliable. Recent advances in solver configuration~\cite{wu2025cubing} and circuit representation learning~\cite{wu2025mgvga,yu2024boolgebra} suggest that  machine learning offers a promising direction for addressing these challenges effectively.

This paper presents a circuit-aware solver configuration framework that employs machine learning for intelligent heuristic selection in PDR-based verification. Our approach addresses the limitations of existing methods through three key innovations:

\begin{itemize}
\item \textbf{Circuit-Aware Feature Extraction:} We develop a comprehensive feature extraction pipeline that captures both graph-level structural properties of And-Inverter Graph (AIG) representations and circuit-level statistical features. This multi-level characterization enables our models to understand the verification complexity and guide parameter selection accordingly.

\item \textbf{Expert-Knowledge-Based Parameter Filtering:} We incorporate human expertise through a systematic constraint-based approach to eliminate invalid parameter combinations based on logical dependencies among PDR options. This filtering mechanism significantly reduces the search space while ensuring that all considered configurations are semantically meaningful.

\item \textbf{GNN-Based PDR Parameter Prediction:} We conduct a comprehensive empirical evaluation of multiple graph neural network architectures, including GraphSAGE~\cite{hamilton2017inductive}, GIN~\cite{xu2018powerful}, GCN~\cite{kipf2016semi}, HOGA~\cite{deng2024less} and GraphSAINT~\cite{zeng2019graphsaint}, to learn the complex mapping between circuit structure and optimal parameters. Through systematic comparison, we identify the most effective architecture for capturing both local node patterns and global graph properties that influence verification performance. Our approach builds upon recent advances in PDR algorithms~\cite{Seufert2022Making,Kori2023Exploiting,Blankestijn2023Incremental} to optimize parameter selection.
\end{itemize}

Our experimental evaluation on a comprehensive benchmark suite demonstrates significant improvements over existing approaches. Our framework increases the number of solved instances by 90.0\% over the default PDR configuration and by 33.3\% over a commonly-used optimized setting. The system successfully identifies circuit-specific optimization patterns and generalizes well across different verification domains.

The main contributions of this work are:
\begin{itemize}
\item A comprehensive machine learning framework for automated PDR parameter tuning that combines circuit feature extraction, parameter filtering, and graph neural network prediction.
\item A systematic parameter filtering mechanism that eliminates invalid configurations based on logical dependencies, significantly reducing the search space.
\item A thorough empirical evaluation of multiple GNN architectures, identifying the most effective approach for circuit-aware solver configuration.
\end{itemize}

The remainder of this paper is organized as follows: Section~\ref{sec:preliminaries} provides background on PDR algorithms and parameter tuning challenges. Section~\ref{sec:method} presents our circuit-aware framework including feature extraction, parameter filtering, and GNN-based prediction. Section~\ref{sec:experiment} describes our experimental setup and evaluation methodology, including presenting and analyzing the experimental results. Section~\ref{sec:conclusion} concludes the paper and discusses future work.

\section{Preliminaries}
\label{sec:preliminaries}
\subsection{Property Directed Reachability (PDR/IC3)}

Property Directed Reachability (PDR)~\cite{bradley2011sat}, also known as IC3, is a SAT-based model checking algorithm that has revolutionized the formal verification of safety properties. Unlike traditional approaches such as Bounded Model Checking (BMC) or fixed-point iteration over symbolic states, PDR constructs a sequence of inductive, relative over-approximations of reachable states in a manner directed by the property being verified. The algorithm maintains a sequence of formulas $F_0, F_1, \ldots, F_k$, where each $F_i$ represents an over-approximation of states reachable in at most $i$ steps from the initial states. PDR incrementally strengthens these formulas by finding and generalizing counterexamples to induction (CTIs). This incremental and goal-oriented approach has made PDR exceptionally effective, particularly in hardware verification~\cite{een2011inductive}. 

The performance of the PDR engine in the \texttt{ABC} verification tool, however, is critically dependent on a set of heuristic parameters that control its core operations. Understanding these parameters requires deep knowledge of PDR's internal mechanisms. Algorithm~\ref{alg:ic3_pdr} presents how heuristic parameters exert their influence in IC3/PDR\footnote{Due to space limit, we refer the readers to the prior works~\cite{bradley2011sat,een2011inductive,Griggio2015} for a more detailed presentation of the algorithm.}. 
During \textbf{Inductive Generalization} (L25-37), several parameters control how counterexamples are generalized. For instance, \texttt{-r} (TwoRound) enables a second round of literal removal for more aggressive generalization. \texttt{-g} (SkipGeneral) bypasses this expensive step entirely. \texttt{-n} (SkipDown) skips the ``down" phase during clause strengthening. The order of generalization is managed by \texttt{-y} (FlopPrio), which uses structural flip-flop priorities to distinguish control logic from datapath, and \texttt{-f} (FlopOrder), which orders flip-flops by these priorities.
For \textbf{Push Propagation} (L11-14), which determines clause propagation strategies, the \texttt{-i} (EagerPush) parameter allows pushing clauses from intermediate frames rather than only from the first frame.
In \textbf{Abstraction \& Refinement} (L4-7), which manages state space reduction, \texttt{-t} (UseAbs) enables abstraction by treating selected latches as pseudo-inputs, while \texttt{-k} (SimpleRefine) employs a simplified refinement strategy when abstraction is used.
These parameters exhibit complex interdependencies that significantly impact PDR's behavior, and their optimal configuration depends heavily on circuit characteristics.


\definecolor{mygood}{RGB}{102, 158, 65}
\definecolor{mybad}{RGB}{242, 98, 105}
\definecolor{myblue}{RGB}{66, 135, 245}
\newcommand{\mygreentext}[1]{\textcolor{mygood}{#1}}
\newcommand{\myredtext}[1]{\textcolor{mybad}{#1}}
\newcommand{\mybluetext}[1]{\textcolor{myblue}{#1}}

\begin{algorithm}
\caption{IC3/PDR with Heuristic Parameters in \texttt{ABC-PDR}}
\label{alg:ic3_pdr}
\begin{algorithmic}[1]
\Require Transition system $(I, T, P)$, safety property $P$
\Ensure \textsc{Safe} or counterexample trace

\State Initialize $F_0 \leftarrow I$, $k \leftarrow 0$
\While{true}
    \State $k \leftarrow k + 1$; $F_k \leftarrow P$

    \State \myredtext{\textbf{// ABSTRACTION \& REFINEMENT ($t$: UseAbs, $k$: SimpleRefine)}}
    \If{$t$} \State Apply latch abstraction \EndIf

    \While{true}
        \State $s \leftarrow$ SAT query: $F_k \land T \land \neg P'$
        \If{$s$ is UNSAT}
            \State \mybluetext{\textbf{// PUSH PROPAGATION ($i$: EagerPush)}}
            \State Push clauses between frames
            \If{$F_j \Leftrightarrow F_{j+1}$ for some $j$} \Return \textsc{Safe} \EndIf
            \State \textbf{break}
        \Else
            \State $c \leftarrow$ \textsc{BlockCTI}$(s, k)$
            \If{$c$ reaches initial states} \Return counterexample \EndIf
        \EndIf
    \EndWhile
\EndWhile

\vspace{0.2cm}
\Procedure{BlockCTI}{$s, \text{level}$}
\State $c \leftarrow$ Extract cube from state $s$
\State \mygreentext{\textbf{// INDUCTIVE GENERALIZATION ($g, r, n, c, y, f$)}}
\If{$\neg g$} \Comment{\texttt{-g}: SkipGeneral}
    \State $c \leftarrow$ \textsc{Generalize}$(c)$ with priority ordering \Comment{\texttt{-y,-f}: FlopPrio, FlopOrder}
    \If{$r$} \State $c \leftarrow$ \textsc{Generalize}$(c)$ \Comment{\texttt{-r}: TwoRounds} \EndIf
    \If{$\neg n$} \State generalize clause with down \Comment{\texttt{-n}: SkipDown} \EndIf
    \If{$c$} \State generalize with CTG \Comment{\texttt{-c}: Ctgs} \EndIf
\EndIf
\State Add $\neg c$ to frames; \Return $c$
\EndProcedure
\end{algorithmic}
\end{algorithm}

\subsection{Graph Neural Networks}

\begin{figure*}[t]
    \centering
    \begin{minipage}[b]{0.58\linewidth}
        \centering
        \includegraphics[width=\linewidth]{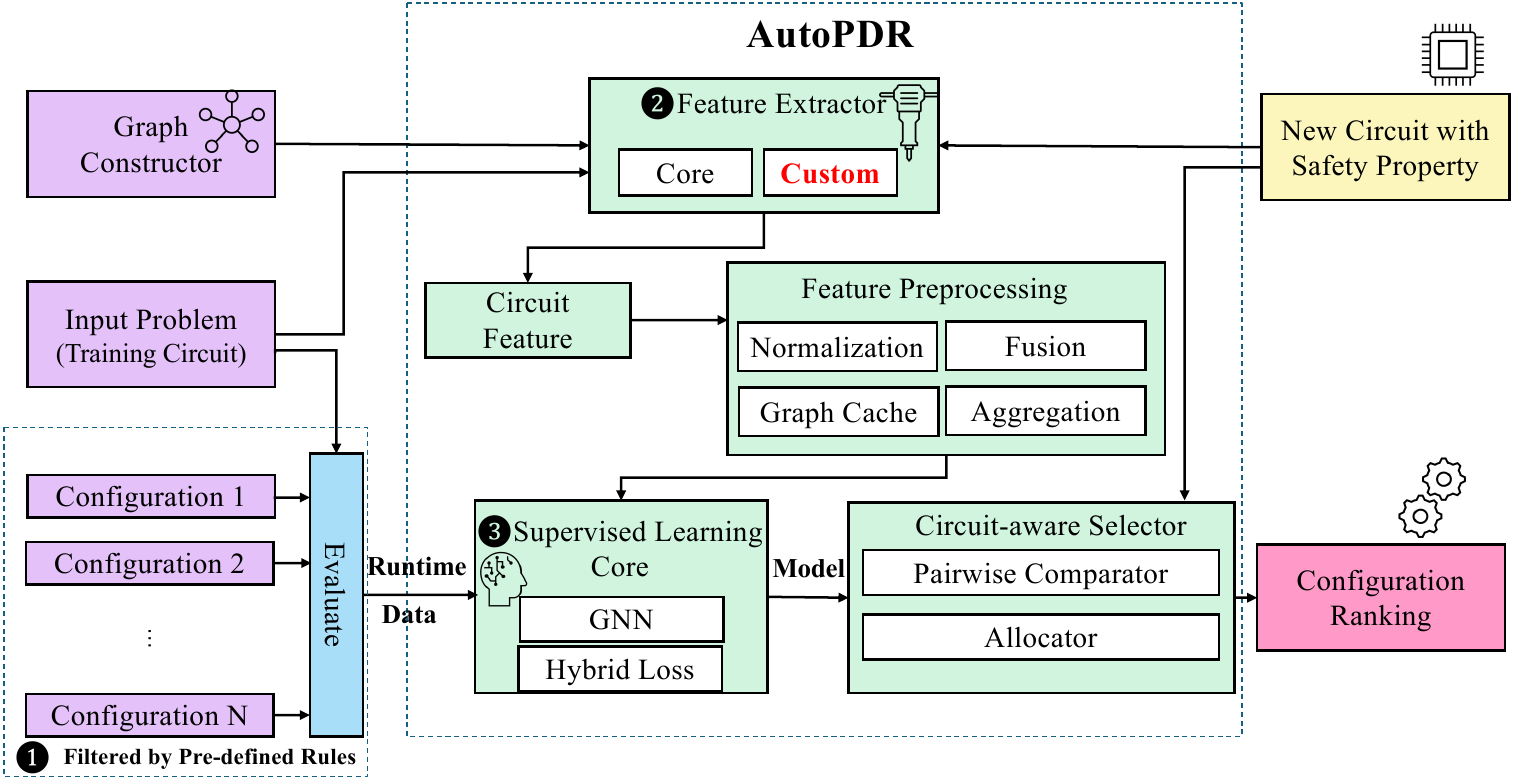}
        \caption{Overview of \AUTOPDR.}
        \label{fig:overview}
    \end{minipage}%
    \hfill
    \begin{minipage}[b]{0.4\linewidth}
        \centering
        \includegraphics[width=\linewidth]{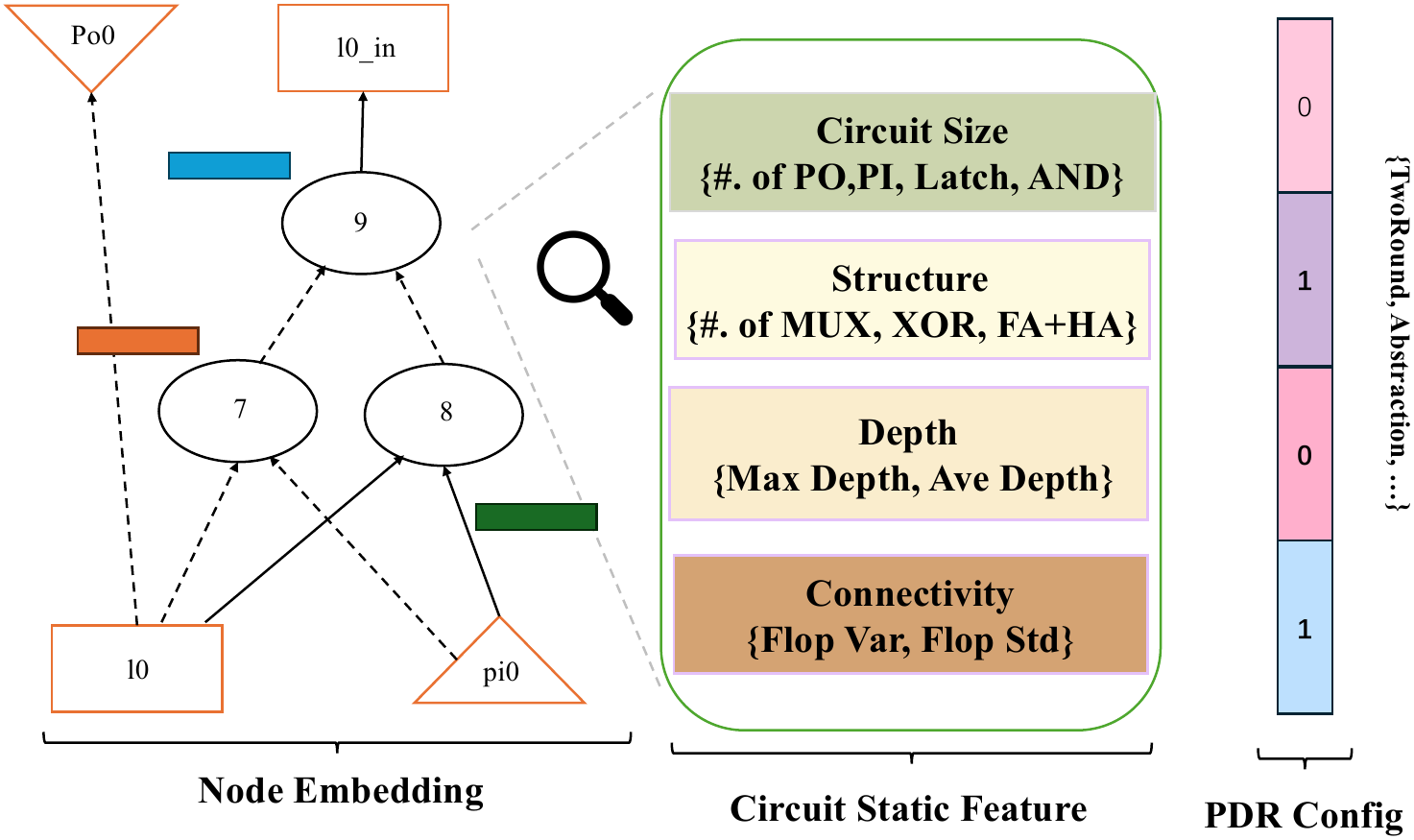}
        \caption{Feature Extraction.}
        \label{fig:feature_extraction}
    \end{minipage}
\end{figure*}

Hardware verification problems are fundamentally graph-structured, with circuits naturally represented as And-Inverter Graphs (AIGs). The structural properties of these graphs—including connectivity patterns, node types, and topological features—are intimately connected to the difficulty of verifying the circuit and the optimal configuration of verification algorithms like PDR. Traditional hand-crafted features often fail to capture the complex structural relationships that influence PDR performance, motivating the use of Graph Neural Networks (GNNs) to automatically learn meaningful circuit representations.

A GNN encodes a graph \( G = (V, E) \) into a more compact representation. Starting with initial node features \( \{ \mathbf{x}^{(0)}_i \mid i \in V \} \), GNNs iteratively update hidden state vectors $\mathbf{x}^{(l)}_i$ using:

\begin{align}
\mathbf{m}^{(l)}_i &= \text{AGGREGATE}^{(l)} \left( \{ \mathbf{x}^{(l-1)}_j \mid j \in \mathcal{N}(i) \} \right), \label{eq:aggregate} \\
\mathbf{x}^{(l)}_i &= \text{COMBINE}^{(l)} \left( \mathbf{m}^{(l)}_i, \mathbf{x}^{(l-1)}_i \right), \label{eq:combine}
\end{align}

where \( \mathcal{N}(i) \) are the one-hop neighbors of node \( i \). The AGGREGATE function (e.g., mean or max pooling) pools neighboring features, and the COMBINE function integrates this with the node's current hidden state vector.

After \( L \) layers, the set of vectors \( \{ \mathbf{x}^{(L)}_i \mid i \in V \} \) captures the structural and feature information of the graph. A readout function then aggregates these vectors into a graph embedding for downstream tasks. For PDR parameter configuration, this enables GNNs to automatically discover which structural patterns in circuit graphs correlate with the effectiveness of different parameter settings, providing a principled approach to solver configuration that goes beyond manual heuristics.


\section{Methodology: Circuit-Aware PDR Solver Configuration}
\label{sec:method}

\subsection{Overview of \AUTOPDR}

\AUTOPDR{} is a machine-learning-driven framework that automatically configures PDR solver parameters based on circuit characteristics. As shown in Figure~\ref{fig:overview}, the framework addresses the fundamental challenge of PDR parameter tuning, leveraging both structural circuit features and graph neural networks to predict optimal parameter configurations for hardware model checking problems.

The core methodology of \AUTOPDR{} consists of three main components: \blackcircled{1} \textbf{Parameter Space Filtering}, which eliminates invalid parameter combinations based on pre-defined rules; \blackcircled{2} \textbf{Circuit Feature Extraction}, which captures key structural and statistical properties from the circuit's graph representation; and \blackcircled{3} \textbf{Learning Circuit-Aware PDR Solver Configuration}, where a Graph Neural Network (GNN) is trained on runtime data to predict the optimal parameter configuration for a new circuit.

Unlike traditional approaches that rely on manual tuning or brute-force search, \AUTOPDR{} exploits the inherent structure of verification problems to guide parameter selection intelligently. The framework is designed to generalize across diverse circuit domains while maintaining computational efficiency during both training and inference phases.

\subsection{Parameter Space Filtering}

\begin{figure}[b]
    \centering
    \includegraphics[width=0.7\linewidth]{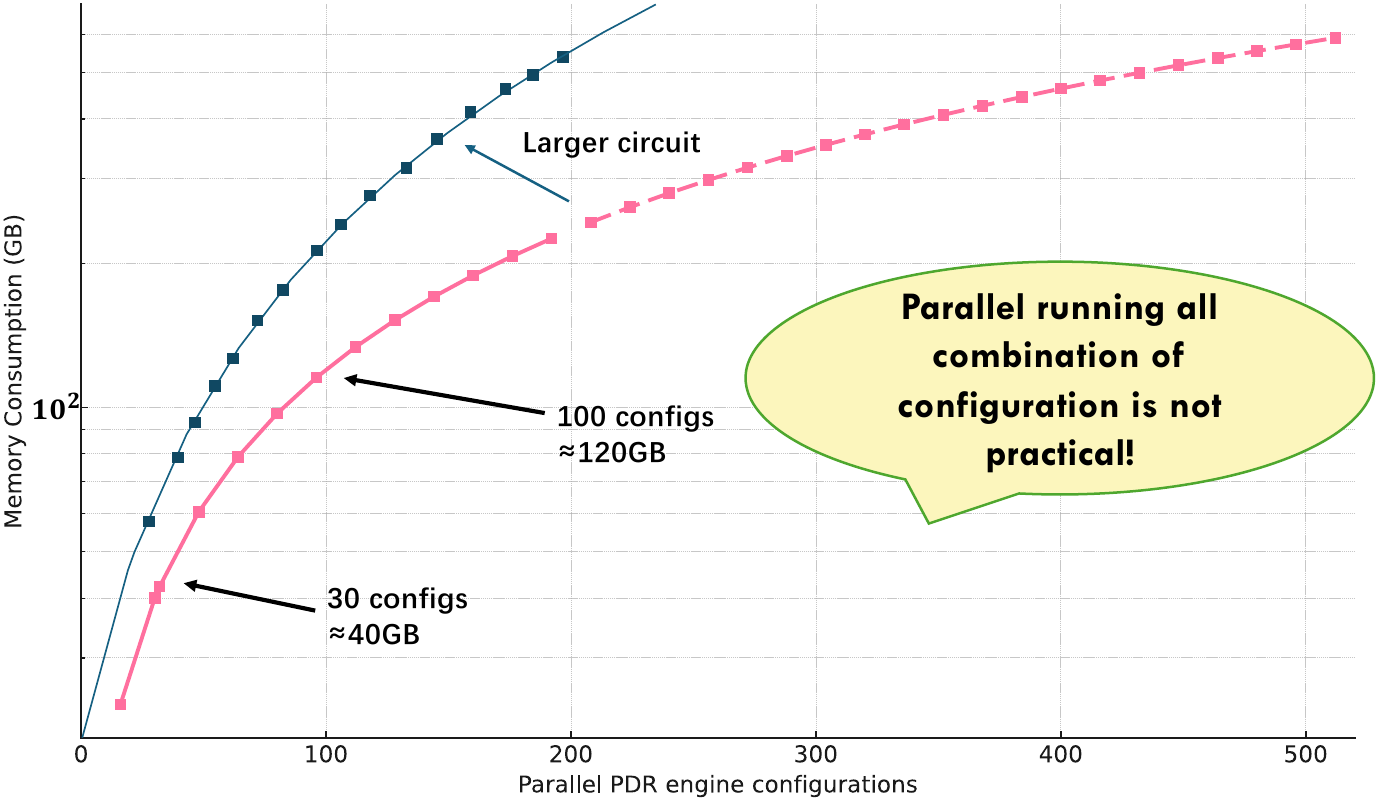}
    \caption{Memory consumption of parallel execution of PDR engines for a small and a large verification problem. The pink line represents the small case \texttt{6s0}  with 157 state bits from the HWMCC13 benchmark. The blue line represents \texttt{ILA\_Flute\_SUB\_problem} with 174263 state bits from the HWMCC24 benchmark.}
    \label{fig:mem_consume}
\end{figure}

The PDR parameter space, with $2^9 = 512$ possible combinations, is too large to explore exhaustively for each input problem. A brute-force approach presents several challenges: 1) Many parameter combinations are logically invalid or counter-productive due to inherent algorithmic dependencies. 2) Evaluating all configurations each time when a new problem is encountered is computationally infeasible, even when parallel execution is employed. As suggested by Figure~\ref{fig:mem_consume}, which shows the memory consumption for one such parallel execution, running all 512 configurations for every circuit would lead to an explosion in memory and CPU core consumption. In general, a smaller and more focused parameter space is crucial for efficient model training and faster prediction.

To address these issues, \AUTOPDR{} incorporates a systematic filtering mechanism to prune the search space. We apply a set of implication rules derived from expert knowledge of the PDR algorithm, as summarized in Table~\ref{tab:parameter_filter_rules}. These rules eliminate invalid or counter-productive configurations by formalizing dependencies between parameters.

\begin{table*}[htbp]
  \centering
  \caption{PDR Parameter Constraint Rules for Search Space Reduction}
  \label{tab:parameter_filter_rules}
  \renewcommand{\arraystretch}{1.3}
  \setlength{\tabcolsep}{6pt}
  \begin{tabular}{@{}p{5cm}p{12cm}@{}}
    \toprule
    \textbf{Parameter Rule} &
    \textbf{Algorithmic Rationale and Implementation Details}\\
    \midrule
     $\mathit{SkipGeneral}{\Rightarrow}\newline\neg\{\mathit{TwoRounds}, \mathit{SkipDown}, \newline\mathit{Ctgs}, \mathit{FlopPrio}, \mathit{FlopOrder}\}$ &
    Skip generalization (\texttt{-g}) is a master switch that bypasses and nullifies several other generalization strategies, including two-round logic (\texttt{-r}), down (\texttt{-n}), CTG handling (\texttt{-c}), and structural ordering (\texttt{-y}, \texttt{-f}). This rule prevents redundant or conflicting operations when high-level generalization is disabled.\\
    \midrule
    $\mathit{FlopOrder} \Rightarrow \mathit{FlopPrio}$ &
    Flop priority (\texttt{-y}) computes structural priorities for flops. These priorities guide two different heuristics: 1) Predecessor generation, by influencing the variable selection in ternary simulation. 2) Inductive generalization, where flop ordering (\texttt{-f}) uses them to sort literals before minimization. For \texttt{-f} to be effective, \texttt{-y} must be enabled to provide meaningful priorities. \\
    \midrule
    $\mathit{Ctgs} \Rightarrow \neg\mathit{SkipDown}$ &
    Counterexample-to-generalization (\texttt{-c}) is a technique for strengthening clauses, which is only effective when the `down' generalization strategy (\texttt{-n}) is active. If `down' is skipped, CTG has no effect and should be disabled.\\
    \midrule
    $\mathit{SimpleRefine} \Rightarrow \mathit{UseAbs}$ &
    Simplified CEGAR refinement (\texttt{-k}) is a subordinate option that modifies the behavior of abstraction-based refinement. Therefore, it only applies when the main abstraction mechanism (\texttt{-t}) is enabled.\\
    \bottomrule
  \end{tabular}
\end{table*}

This rule-based filtering significantly reduces the search space to 114 valid configurations, an elimination of approximately 78\% of the original space. The resulted set of configurations is not only logically sound but also computationally manageable, forming a diverse and effective basis for training our prediction model.

\subsection{Circuit Feature Extraction and Graph Representation}

\AUTOPDR{} leverages a dual-representation approach to obtain a comprehensive understanding of each circuit, as illustrated in Figure~\ref{fig:feature_extraction}. We extract high-level static features from the entire circuit structure to provide a holistic, quantitative summary. Simultaneously, we generate a detailed graph representation that is optimized for processing by a Graph Neural Network (GNN), allowing the model to learn complex topological patterns.

\subsubsection{Graph Representation Preprocessing via Cone of Influence}

To prepare a circuit for the GNN, we first convert its And-Inverter Graph (AIG) into a graph data structure using the \texttt{\&edgelist} command, which generates both the edge list and an initial set of node features. However, industrial-scale circuits can result in massive graphs that are computationally expensive to process. To address this, we apply a \textit{cone of influence} (COI) reduction as a crucial preprocessing step. This optimization prunes logic that does not affect the primary outputs, significantly reducing the graph's size and complexity.

Crucially, this COI reduction is applied \textit{only} to generate the graph for the GNN. The static features, described next, are extracted from the original, unmodified circuit to ensure their accuracy. This dual approach provides the best of both worlds: complete, accurate features for high-level analysis and an efficient, optimized graph for deep learning.

\begin{figure}[h]
    \centering
    \begin{minipage}[t]{0.48\linewidth}
        \centering
        \includegraphics[width=\linewidth]{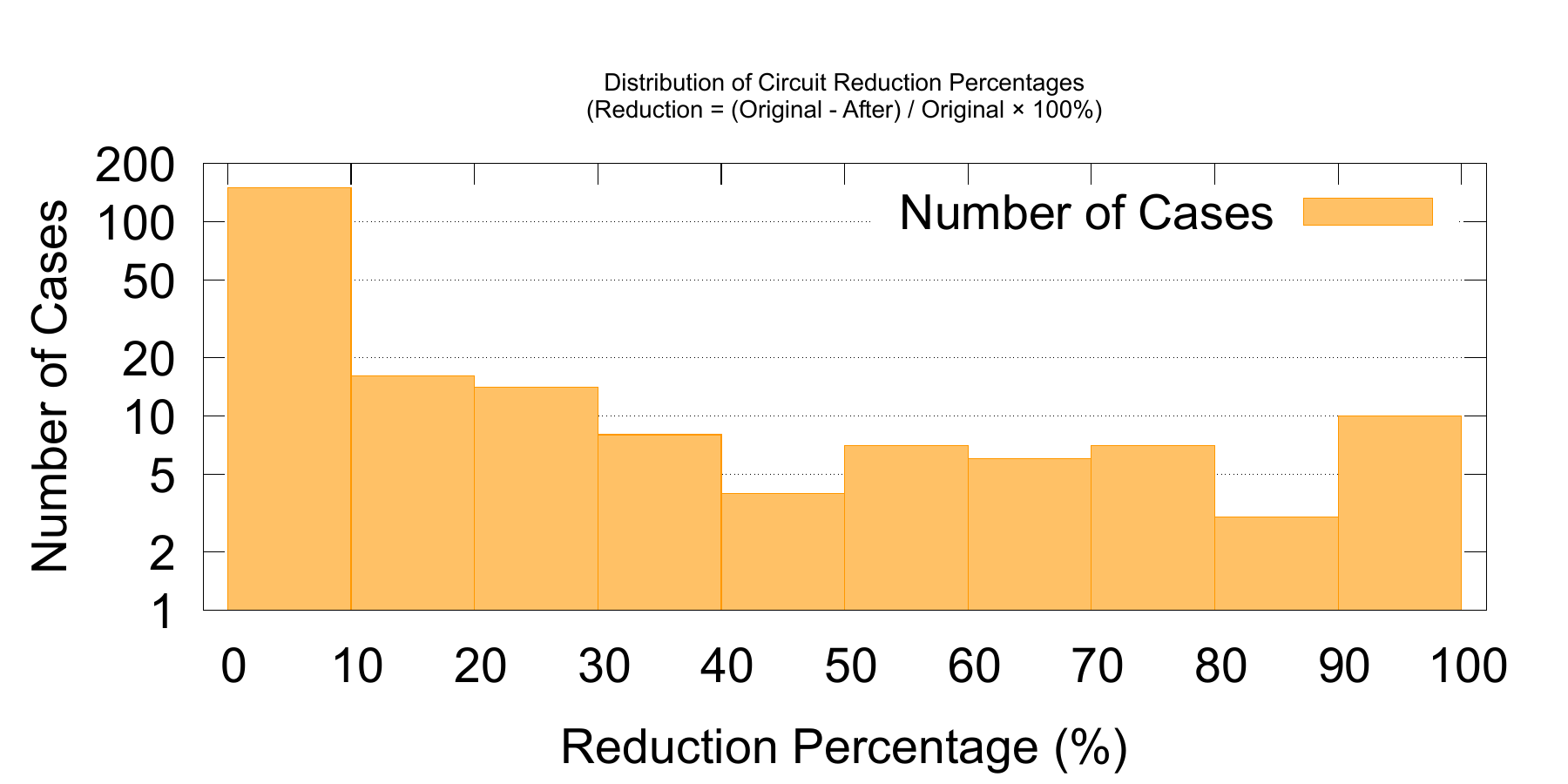}
        \caption{Distribution of COI reduction percentages across 225 benchmarks from HWMCC'13.}
        \label{fig:cone_reduction}
    \end{minipage}
    \hfill
    \begin{minipage}[t]{0.48\linewidth}
        \centering
        \includegraphics[width=\linewidth]{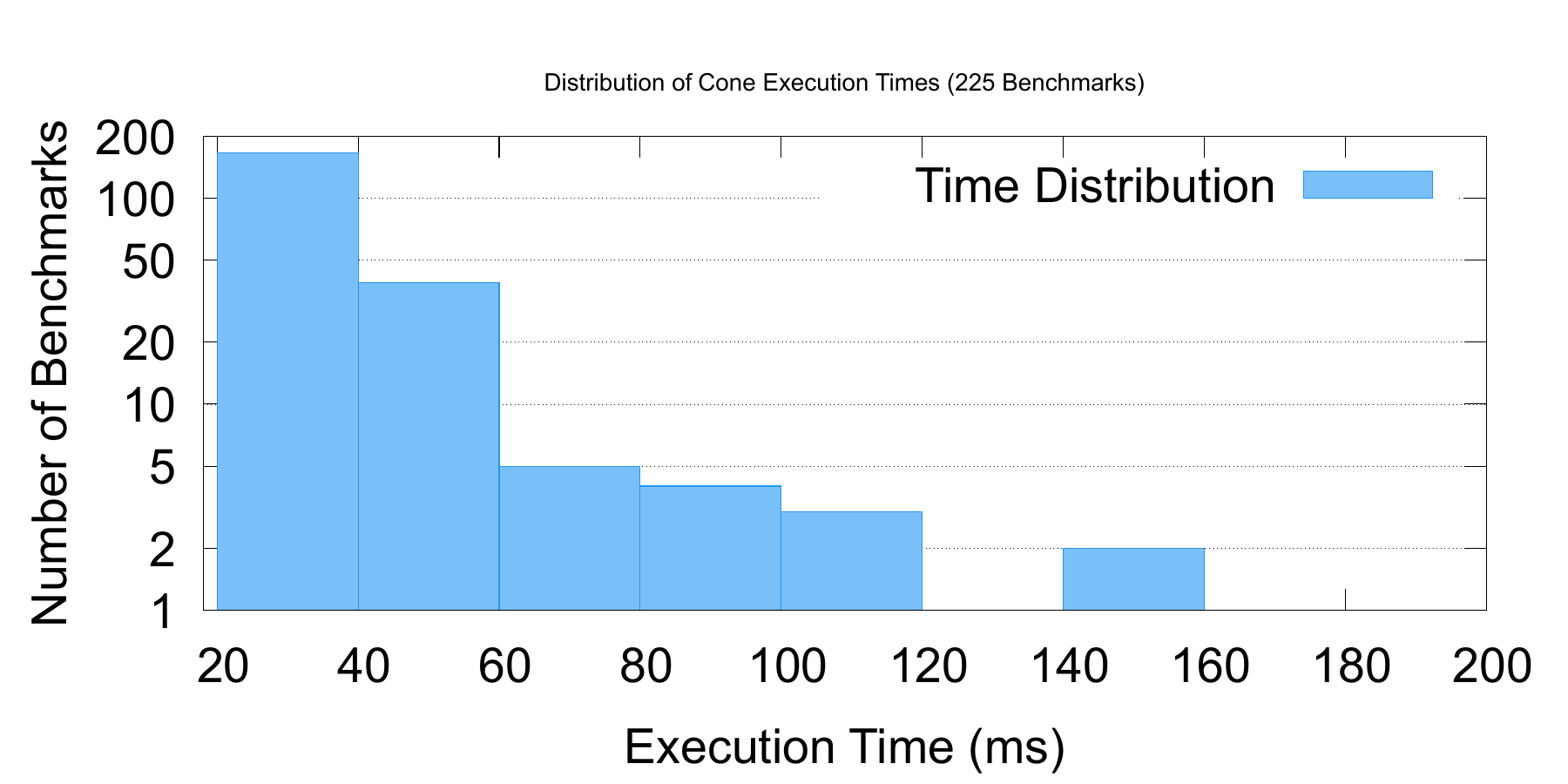}
        \caption{Distribution of COI execution times across the same 225 benchmarks.}
        \label{fig:cone_time_distribution}
    \end{minipage}
\end{figure}

To validate this preprocessing step, we conducted a preliminary analysis on the single-property AIGER benchmarks from HWMCC'13. As shown in Figure~\ref{fig:cone_reduction} and Figure~\ref{fig:cone_time_distribution}, this study confirmed that COI reduction is both effective and efficient, eliminating an average of 16.49\% of AND gates and completing in under 100ms for most instances, thus justifying its inclusion in our main experimental pipeline.

\subsubsection{Static Feature Extraction}
It captures both the structural complexity and statistical properties of the circuit. As depicted in Figure~\ref{fig:feature_extraction}, it operates on the original AIG and encompasses the following four critical categories, chosen for its direct relevance to the PDR parameters we aim to predict:

\subsubsection*{\textbf{Topological Scale and Complexity}} This category quantifies the fundamental scale of the verification problem. We measure the number of primary inputs (PI), primary outputs (PO), latches (flip-flops), and AND gates. These metrics are critical for deciding on high-level trade-offs between proof search time and generalization effort. For instance, for exceptionally large circuits, the cost of the main generalization procedure can become prohibitively high, suggesting that it should be skipped entirely (\texttt{SkipGeneral}). Furthermore, a high latch count indicates a large state space, which is the primary motivation for enabling abstraction-based techniques (\texttt{UseAbs}).

\subsubsection*{\textbf{High-Level Functional Composition}} This category identifies high-level functional patterns that are crucial for guiding PDR's generalization strategies. We count the number of multiplexers (MUX), XOR gates, and arithmetic units (Full and Half Adders). The presence of these functional blocks is a strong indicator that the solver may encounter complex Counterexamples-To-Generalization (CTGs) that standard literal-dropping cannot easily resolve. The ``down" phase of clause strengthening is a powerful technique for handling such CTGs. Therefore, a high count of these functional structures suggests that disabling this phase (\texttt{SkipDown}) would likely be detrimental to performance.

\subsubsection*{\textbf{Logic Depth and Propagation Characteristics}} This category characterizes the circuit's combinational logic complexity through its maximum and average depth. These metrics are particularly relevant for PDR's frame-based reasoning. In circuits with deep logic, reaching a fixed-point inductive invariant can be slow. Enabling ``eager" clause pushing (\texttt{EagerPush}) allows the solver to start propagating clauses from intermediate frames, potentially accelerating convergence. Additionally, deeper logic can complicate the abstraction-refinement loop; these features can inform the choice between standard and simplified refinement strategies (\texttt{SimpleRefine}) when abstraction (\texttt{UseAbs}) is active.

\subsubsection*{\textbf{State-Space Connectivity and Influence}} This category analyzes signal propagation patterns by computing the variance and standard deviation of flip-flop fanouts. These statistics capture the distribution of state variable dependencies. A high variance often indicates a clear separation between control logic (low fanout flops) and datapath logic (high fanout flops). This structural information provides a direct, quantitative basis for enabling structural flip-flop priorities (\texttt{FlopPrio}) and ordering the flip-flops by this priority during generalization (\texttt{FlopOrder}).

The extracted features are systematically normalized and encoded into an 11-dimensional feature vector $\mathbf{f}_{\text{circuit}} \in \mathbb{R}^{11}$ through a standardized transformation process. This encoding ensures consistent scaling across diverse circuit domains while preserving the essential structural characteristics that govern PDR solver performance, enabling effective machine learning-based parameter prediction.

\subsection{Learning Circuit-Aware PDR Solver Configuration}

\subsubsection{Training Data Selection}
The quality and composition of the data are fundamental to the model's predictive power. To ensure the model learns from meaningful and challenging instances, we first curated a dataset from the Hardware Model Checking Competition (HWMCC) benchmarks, spanning from 2007 to 2024. We specifically selected non-trivial cases where the solving time exceeded 60 seconds, filtering out instances that are too easy to provide a strong learning signal. This selection process resulted in a focused dataset comprising approximately 5\% of the total available benchmarks. This dataset was then randomly partitioned into training (80\%), validation (10\%), and test (10\%) sets to ensure a rigorous and unbiased evaluation of the model's performance.

\subsubsection{Empirical Evaluation of GNN Architecture}
After establishing the circuit-aware hybrid representation, we design a GNN-based model to predict PDR solver runtime based on hardware circuit characteristics. We evaluated several prominent GNN models and use Kendall's $\tau$ and Spearman's $\rho$ as evaluation metrics, which measure the ordinal association between predicted and actual rankings of PDR parameter configurations for each hardware circuit in model checking tasks:
\begin{equation}
\tau = \frac{N_c - N_d}{n(n-1)/2}, \quad \rho = 1 - \frac{6 \sum d_i^2}{n(n^2 - 1)}
\end{equation}
where $N_c$ and $N_d$ are concordant and discordant pairs, and $d_i$ is the rank difference for each observation.

Table~\ref{tab:gnn_performance} shows the performance of various GNN encoders on the held-out test set. \textbf{GraphSAGE} emerges as the top-performing architecture, achieving the highest Kendall's $\tau$ (0.508) and Spearman's $\rho$ (0.683). Its inductive learning capability, which allows it to generalize to unseen hardware circuit graphs, and its neighborhood aggregation mechanism are particularly well-suited for capturing the circuit-specific structural and topological features that influence PDR solver performance on different hardware verification problems.

\begin{table}[htbp]
\centering
\caption{Performance comparison for GNN encoders on the test set. BiGraphSAGE uses two separate SAGE modules to aggregate from predecessors and successors in directed graphs and then combines their embeddings.}
\label{tab:gnn_performance}
\begin{tabular}{@{}lcc@{}}
\toprule
\textbf{Encoder} & \textbf{Kendall's $\tau$} & \textbf{Spearman's $\rho$} \\
\midrule
\textbf{GraphSAGE~\cite{hamilton2017inductive}} & \textbf{0.508} & \textbf{0.683} \\
Bi-GraphSAGE & 0.432 & 0.640 \\
GraphSAINT~\cite{zeng2019graphsaint} & 0.446 & 0.594 \\
GCN~\cite{kipf2016semi} & 0.348 & 0.512 \\
GIN~\cite{xu2018powerful} & 0.390 & 0.520 \\
HOGA~\cite{deng2024less} & 0.484 & 0.583 \\
\bottomrule
\end{tabular}
\end{table}

\begin{figure}[htbp]
  \centering
  \includegraphics[width=0.85\linewidth]{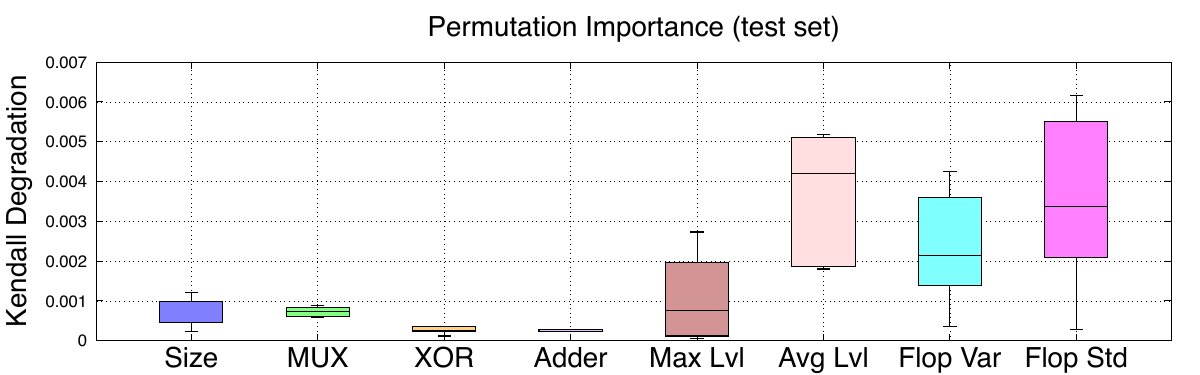}
  \caption{Permutation test for feature importance. The y-axis shows the degradation in Kendall's $\tau$ when a feature is permuted; higher values indicate greater importance.}
  \label{fig:feature-importance}
\end{figure}

The model first computes node-level embeddings using the GraphSAGE encoder, which are then aggregated into a single circuit topology embedding, $\mathbf{h}_{\text{circuit}}$, via a global mean pooling operation. This embedding is then concatenated with the verification-oriented feature vector $\mathbf{f}_{\text{circuit}}$ and the target PDR parameter vector $\mathbf{p}_{\text{pdr}}$:
\begin{equation}
\mathbf{h}_{\text{combined}} = \text{concat}(\mathbf{h}_{\text{circuit}}, \mathbf{f}_{\text{circuit}}, \mathbf{p}_{\text{pdr}})
\end{equation}
This combined vector, which encapsulates hardware circuit features and model checking information, is fed into a multi-layer perceptron (MLP)  to predict the final PDR solver runtime for the specific circuit-parameter combination. Figure~\ref{fig:feature-importance} shows the relative importance of the verification-oriented circuit features, confirming that our selected hardware circuit properties are highly relevant to the PDR solver configuration prediction task.

Figure~\ref{fig:feature-importance} illustrates the relative importance of different verification-oriented hardware circuit features, as determined by a permutation test. The test measures the degradation in Kendall's $\tau$ when the values of a single circuit feature are randomly shuffled. A larger degradation indicates higher importance for hardware circuit-specific PDR solver configuration prediction. The results show that \textbf{Flop Std} (standard deviation of flip-flop fanout) and \textbf{Avg Lvl} (average logic level) are the most influential hardware circuit characteristics. This aligns with our model checking domain knowledge, as flip-flop connectivity patterns and combinational logic depth are known to critically impact PDR solver behavior on different hardware verification problems.

\begin{figure*}[htbp]
  \centering
  \begin{subfigure}[b]{0.3\textwidth}
    \centering
    \includegraphics[width=\linewidth]{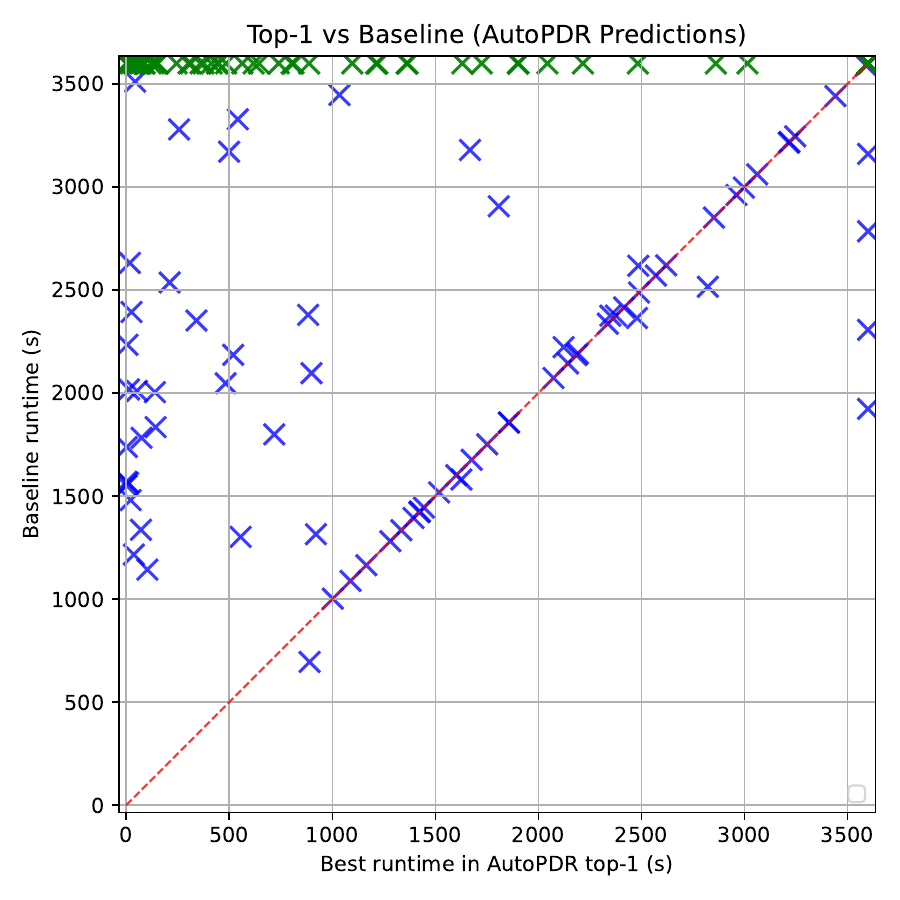}
    \caption{Top-1 predicted config. vs baseline \textit{pdr}}
    \label{fig:top1_predict}
  \end{subfigure}
  \hfill
  \begin{subfigure}[b]{0.3\textwidth}
    \centering
    \includegraphics[width=\linewidth]{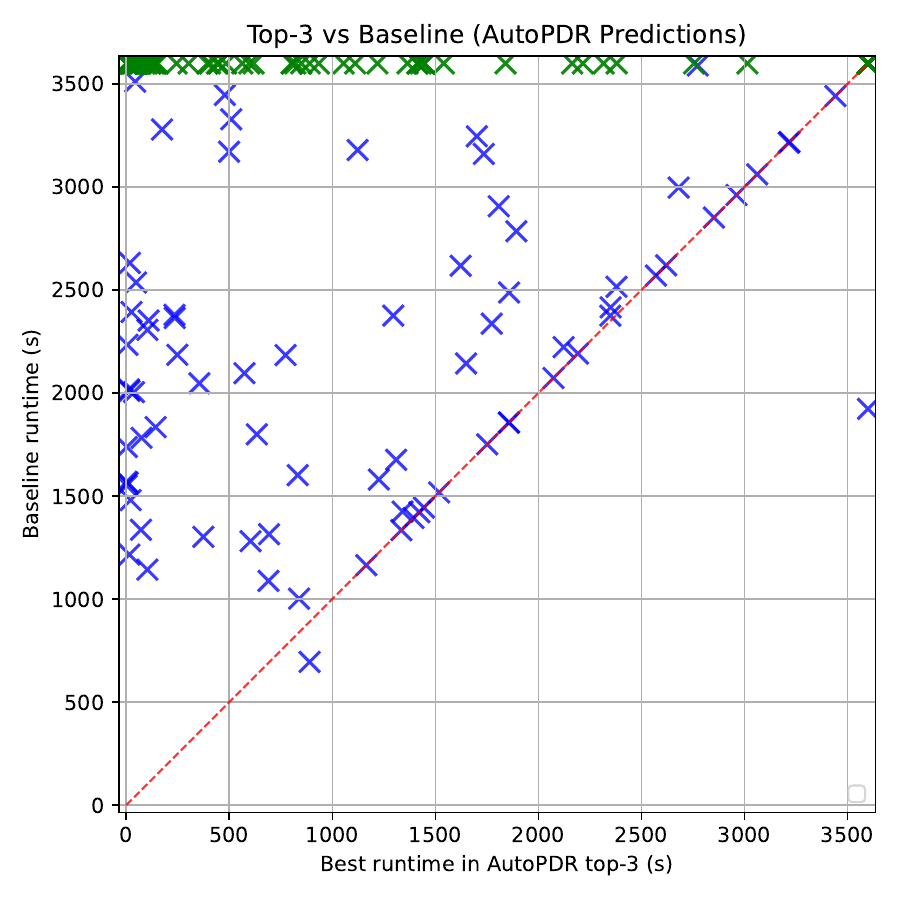}
    \caption{Top-3 predicted config. vs baseline \textit{pdr}}
    \label{fig:top3_predict}
  \end{subfigure}
  \hfill
  \begin{subfigure}[b]{0.3\textwidth}
    \centering
    \includegraphics[width=\linewidth]{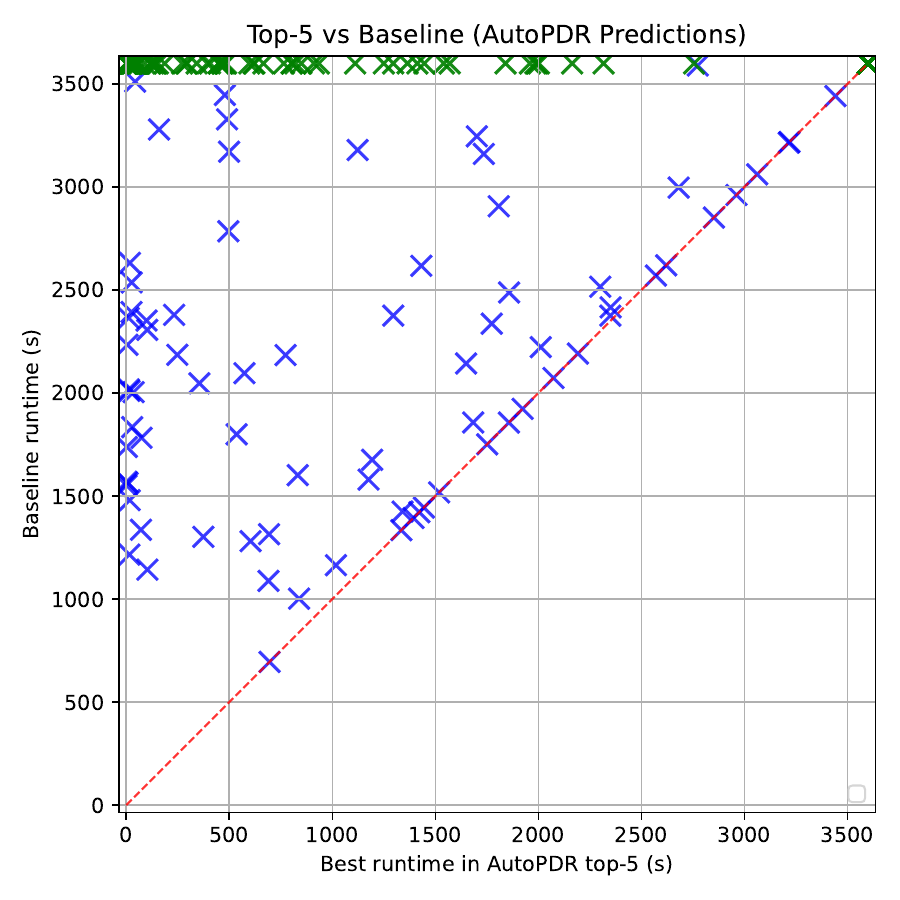}
    \caption{Top-5 predicted config. vs baseline \textit{pdr}}
    \label{fig:top5_predict}
  \end{subfigure}
  \caption{Performance comparison between \AUTOPDR's Top-k predicted configurations and the default configuration. When $k>1$, the shortest runtime among the Top-k configurations is used. Points above the diagonal line indicate cases where our predicted configurations outperform the baseline \textit{pdr}.}
  \label{fig:topk_predictions}
\end{figure*}

\subsubsection{Rank-Aware Hybrid Loss Function}
A key insight of our work is that precisely predicting the absolute runtime of a PDR configuration on a specific circuit is a secondary goal. The primary objective for practical application is to \textbf{correctly rank the configurations} for each circuit to identify the best one for that particular circuit. A standard Mean Squared Error (MSE) loss is insufficient as it equally penalizes all prediction errors and does not explicitly optimize for circuit-specific ranking quality.

To address this, we introduce a \textbf{Rank-Aware Hybrid Loss Function} that aligns the training objective with the ultimate goal of ranking. This loss function combines a regression component with a pairwise ranking component:
\begin{equation}
\mathcal{L}_{\text{total}} = \alpha \cdot \mathcal{L}_{\text{MSE}} + (1-\alpha) \cdot \mathcal{L}_{\text{ranking}}
\label{eq:hybrid_loss}
\end{equation}
where $\alpha$ is a balancing hyperparameter, which we set to 0.3.

The regression component, $\mathcal{L}_{\text{MSE}}$, is computed on the log-transformed runtimes to handle their wide dynamic range and focus on relative error. The ranking component, $\mathcal{L}_{\text{ranking}}$, is a pairwise hinge loss that directly penalizes incorrectly ordered pairs of configurations. For any two configurations $i$ and $j$ in a batch, where the true runtime $t_i < t_j$, the loss is formulated as:
\begin{equation}
\mathcal{L}_{\text{ranking}} = \frac{1}{|\mathcal{P}|} \sum_{(i,j) \in \mathcal{P}} \max(0, m - (\hat{t}_j - \hat{t}_i))
\label{eq:ranking_loss}
\end{equation}
where $\mathcal{P}$ is the set of all pairs where $t_i < t_j$, $\hat{t}$ are the predicted runtimes, and $m$ is a margin hyperparameter. This loss encourages the model to predict a lower runtime for configuration $i$ than for $j$, thus preserving their relative order.

By optimizing this hybrid objective, \AUTOPDR{} learns not only to estimate runtimes but, more importantly, to discern the subtle performance differences between parameter sets for specific circuits, making it highly effective at its core task of finding the optimal configuration for each individual circuit.


\section{Experiments}
\label{sec:experiment}


\subsection{Experimental Setup}
Our experimental setup is designed to rigorously evaluate the performance and generalization capabilities of \AUTOPDR. The framework is implemented in Python using PyTorch and PyTorch Geometric, with circuit feature extraction and graph construction accelerated via a Rust interface to \texttt{ABC}~\cite{brayton2010abc}. All experiments were conducted on a Ubuntu 20.04.4 LTS server with Intel Xeon Platinum 8375C CPUs, an NVIDIA 3090 GPU, and 128GB of memory, with a one-hour wall-clock time limit per task.

To assess the generalization capability of our method, we carefully curated a challenging evaluation benchmark of 192 circuits from HWMCC07 to HWMCC24 under the following criteria:
\begin{itemize}
    \item \textbf{Nontrivial and unsolved problems:} To test the framework's ability to handle difficult cases, we included instances where the baseline \textit{pdr} configuration exceeded a 900-second solve time, many of which were previously unsolved.
    \item \textbf{Circuits with unseen properties:} To evaluate generalization across different verification tasks on the same circuit, we included instances that may share the name with a circuit in the training set but target a different property.
    \item \textbf{Unseen circuits:} To assess broader generalization, we included circuits that may share the same family name as the training data (e.g., `6s', `intel') but with distinct circuit structures.
\end{itemize}
These selection criteria ensure that our evaluation is conducted on a diverse and challenging set of problems not seen during training. 

\begin{table}[htbp]
  \caption{Performance Comparison with Baseline Configurations}
  \label{tab:baseline_comparison}
  \centering
  \renewcommand{\arraystretch}{1.1}
  \setlength{\tabcolsep}{4pt}
  \footnotesize
  \definecolor{lightblue}{RGB}{173, 216, 230}
  \begin{tabular}{@{}lccccl@{}}
    \toprule
    \textbf{Method} & \textbf{Safe} & \textbf{Unsafe} & \textbf{Timeout} & \textbf{Solved} & \textbf{Improvement} \\
    \midrule
    Baseline (\textit{pdr}) & 65 & 15 & 112 & 80 & Baseline \\
    Baseline (\textit{pdr-nct}) & 82 & 32 & 78 & 114 & +34 (42.5\%) \\
    \AUTOPDR Top-1 & \colorbox{lightblue}{\textbf{94}} & 27 & 71 & \colorbox{lightblue}{\textbf{121}} & \colorbox{lightblue}{\textbf{+41 (51.2\%)}} \\
    \AUTOPDR Top-3 & \colorbox{lightblue}{\textbf{109}} & 31 & 52 & \colorbox{lightblue}{\textbf{140}} & \colorbox{lightblue}{\textbf{+60 (75.0\%)}} \\
    \AUTOPDR Top-5 & \colorbox{lightblue}{\textbf{117}} & \colorbox{lightblue}{\textbf{35}} & 40 & \colorbox{lightblue}{\textbf{152}} & \colorbox{lightblue}{\textbf{+72 (90.0\%)}} \\
    \bottomrule
  \end{tabular}
\end{table}

\subsection{Performance Analysis}
We analyze the performance of \AUTOPDR by comparing its Top-k predicted configurations against two baselines: the default \textit{pdr} configuration and a commonly-used optimized setting, \textit{pdr-nct}.

\Cref{tab:baseline_comparison} summarizes the results, showing that \AUTOPDR significantly improves solving capability. The Top-5 configurations achieve the highest performance, solving 152 out of 192 cases (79.2\% success rate) compared to the default's 80 solved cases (41.7\%). This represents a 90.0\% increase in solved instances. The performance progressively improves from Top-1 (121 solved) to Top-3 (140 solved) and Top-5 (152 solved), demonstrating that providing multiple diverse configurations enhances the likelihood of finding an optimal parameter set.

\Cref{fig:topk_predictions} provides a visual comparison between \AUTOPDR's Top-k predictions and the default baseline. Each scatter plot compares the runtime of a predicted configuration (x-axis) against the baseline runtime (y-axis). Points above the diagonal indicate instances where \AUTOPDR's prediction is superior. A significant number of points fall in this region, particularly for Top-3 and Top-5 predictions, highlighting consistent performance gains. The plots also distinguish between cases where the baseline solved the instance (blue) and where it timed out (green), underscoring our approach's effectiveness on challenging verification tasks.

\section{Conclusion}
\label{sec:conclusion}

This paper introduced \AUTOPDR{}, a novel machine learning framework for automated PDR parameter tuning. By combining systematic parameter filtering with a GNN-based model that learns from both static and graph-based circuit features, our approach effectively navigates the complex PDR parameter space. Our experimental evaluation demonstrated that \AUTOPDR{} significantly outperforms default configurations, increasing the number of solved instances by 90.0\%. The framework successfully identifies circuit-specific parameter patterns and generalizes to unseen problems, making it a valuable tool for automated formal verification workflows. Future work will focus on extending the framework to a wider range of parameters and exploring dynamic, reinforcement learning-based adaptation.






\bibliographystyle{IEEEtran}
\bibliography{bib/references}

@inproceedings{bradley2011sat,
    title     = {{SAT}-based model checking without unrolling},
    author    = {Bradley, AR},
    year      = {2011},
    booktitle = {International Workshop on Verification, Model Checking, and Abstract Interpretation},
    pages     = {70-87},
    publisher = {Springer Berlin Heidelberg}
}

@inproceedings{een2011inductive,
    author    = {Niklas E{\'e}n and Alan Mishchenko and Robert Brayton},
    title     = {Efficient Implementation of Property Directed Reachability},
    booktitle = {Formal Methods in Computer-Aided Design},
    year      = {2011},
    pages     = {125--134}
}

@inproceedings{hoder2011generalized,
    author    = {Krystof Hoder and Nikolaj Bj{\o}rner},
    title     = {Generalized Property Directed Reachability},
    booktitle = {International Conference on Computer Aided Verification},
    year      = {2012},
    pages     = {157--171},
    publisher = {Springer}
}

@inproceedings{kipf2016semi,
    author    = {Thomas N. Kipf and Max Welling},
    title     = {Semi-Supervised Classification with Graph Convolutional Networks},
    booktitle = {International Conference on Learning Representations},
    year      = {2017}
}

@inproceedings{hamilton2017inductive,
    author    = {Will Hamilton and Zhitao Ying and Jure Leskovec},
    title     = {Inductive Representation Learning on Large Graphs},
    booktitle = {Advances in Neural Information Processing Systems},
    year      = {2017}
}

@inproceedings{Seufert2022Making,
    author    = {Seufert, T. and Scholl, C. and Chandrasekharan, A. and Reimer, S. and Welp, T.},
    title     = {Making {PROGRESS} in property directed reachability},
    booktitle = {International Conference on Verification, Model Checking, and Abstract Interpretation},
    year      = {2022},
    pages     = {355-377},
    publisher = {Springer International Publishing},
    address   = {Cham}
}

@inproceedings{Kori2023Exploiting,
    author    = {Kori, Mayuko and Ascari, Flavio and Bonchi, Filippo and Bruni, Roberto and Gori, Roberta and Hasuo, Ichiro},
    title     = {Exploiting adjoints in property directed reachability analysis},
    booktitle = {International Conference on Computer Aided Verification},
    year      = {2023},
    pages     = {41-63},
    publisher = {Springer Nature Switzerland},
    address   = {Cham}
}

@inproceedings{Blankestijn2023Incremental,
    author    = {Blankestijn, Max and Laarman, Alfons},
    title     = {Incremental property directed reachability},
    booktitle = {International Conference on Formal Engineering Methods},
    year      = {2023},
    pages     = {208-227},
    publisher = {Springer Nature Singapore},
    address   = {Singapore}
}

@article{wu2025mgvga,
    author  = {Haoyuan Wu and others},
    title   = {Circuit Representation Learning with Masked Gate Modeling and {Verilog-AIG} Alignment},
    journal = {arXiv preprint arXiv:2502.12732},
    year    = {2025}
}

@article{yu2024boolgebra,
    author  = {Cunxi Yu and others},
    title   = {{BoolGebra}: Attributed Graph-learning for Boolean Algebraic Manipulation},
    journal = {arXiv preprint arXiv:2401.10753},
    year    = {2024}
}

@article{wu2025cubing,
    author  = {Haoze Wu and others},
    title   = {Cubing for Tuning},
    journal = {arXiv preprint arXiv:2504.19039},
    year    = {2025}
}

@article{hutter2016cssc,
    author  = {Frank Hutter and Marius Lindauer and Adrian Balint and Sam Bayless and Holger Hoos and Kevin Leyton-Brown},
    title   = {The Configurable SAT Solver Challenge (CSSC)},
    journal = {Artificial Intelligence},
    year    = {2016}
}

@article{leventi2021ml_sat,
    author  = {A.-M. Leventi-Peetz and J.-V. Peetz and M. Rohde},
    title   = {ML Supported Predictions for SAT Solvers Performance},
    journal = {arXiv preprint arXiv:2112.09438},
    year    = {2021}
}

@article{Griggio2015,
    author  = {Griggio, Alberto and Roveri, Marco},
    title   = {Comparing different variants of the {IC3} algorithm for hardware model checking},
    journal = {IEEE Transactions on Computer-Aided Design of Integrated Circuits and Systems},
    volume  = {35},
    number  = {6},
    year    = {2015},
    pages   = {1026-1039}
}

@article{Cabodi2017,
    author  = {Cabodi, Gianpiero and Camurati, P. E. and Mishchenko, A. and Palena, M. and Pasini, P.},
    title   = {{SAT} solver management strategies in {IC3}: an experimental approach},
    journal = {Formal Methods in System Design},
    year    = {2017},
    volume  = {50},
    pages   = {39-74}
}

@inbook{Hoos2012,
    author    = {Hoos, Holger H},
    title     = {Automated algorithm configuration and parameter tuning},
    year      = {2012},
    booktitle = {Autonomous search},
    pages     = {37-71},
    publisher = {Springer Berlin Heidelberg},
    address   = {Berlin, Heidelberg}
}

@inproceedings{brayton2010abc,
  title={{ABC}: An academic industrial-strength verification tool},
  author={Brayton, Robert and Mishchenko, Alan},
  booktitle={Computer Aided Verification: 22nd International Conference, CAV 2010, Edinburgh, UK, July 15-19, 2010. Proceedings 22},
  pages={24--40},
  year={2010},
  organization={Springer}
}

@inproceedings{deng2024less,
  title={Less is more: Hop-wise graph attention for scalable and generalizable learning on circuits},
  author={Deng, Chenhui and Yue, Zichao and Yu, Cunxi and Sarar, Gokce and Carey, Ryan and Jain, Rajeev and Zhang, Zhiru},
  booktitle={Proceedings of the 61st ACM/IEEE Design Automation Conference},
  pages={1--6},
  year={2024}
}

@article{xu2018powerful,
  title={How powerful are graph neural networks?},
  author={Xu, Keyulu and Hu, Weihua and Leskovec, Jure and Jegelka, Stefanie},
  journal={arXiv preprint arXiv:1810.00826},
  year={2018}
}

@article{zeng2019graphsaint,
  title={Graphsaint: Graph sampling based inductive learning method},
  author={Zeng, Hanqing and Zhou, Hongkuan and Srivastava, Ajitesh and Kannan, Rajgopal and Prasanna, Viktor},
  journal={arXiv preprint arXiv:1907.04931},
  year={2019}
}


\end{document}